# Performance of a 4 Kelvin pulse-tube cooled cryostat with dc SQUID amplifiers for bolometric detector testing

D. Barron, M. Atlas, B. Keating, R. Quillin, N. Stebor, B. Wilson

Center for Astrophysics and Space Sciences, Univ. of California San Diego
La Jolla, CA USA 92093-0424

**ABSTRACT**

The latest generation of cosmic microwave background (CMB) telescopes is searching for the undetected faint signature of gravitational waves from inflation in the polarized signal of the CMB. To achieve the unprecedented levels of sensitivity required, these experiments use arrays of superconducting Transition Edge Sensor (TES) bolometers that are cooled to sub-Kelvin temperatures for photon-noise limited performance. These TES detectors are read out using low-noise SQUID amplifiers. To rapidly test these detectors and similar devices in a laboratory setting, we constructed a cryogenic refrigeration chain consisting of a commercial two-stage pulse-tube cooler, with a base temperature of 3 K, and a closed-cycle 3He/4He/3He sorption cooler, with a base temperature of 220 mK. A commercial dc SQUID system, with sensors cooled to 4 K, was used as a highly-sensitive cryogenic ammeter. Due to the extreme sensitivity of SQUIDs to changing magnetic fields, there are several challenges involving cooling them with pulse-tube coolers. Here we describe the successful design and implementation of measures to reduce the vibration, electromagnetic interference, and other potential sources of noise associated with pulse-tube coolers.

**INTRODUCTION**

Commercial cryocoolers become an attractive alternative to liquid cryogen cooling for experimental studies requiring temperatures below 4 K. As the reliability, cost, and cooling power of cryocoolers have improved, the main drawbacks remain the electromagnetic interference (EMI), mechanical vibration, and temperature fluctuations inherent to cryocoolers.[1] These effects can have a significant impact on sensitive measurements, especially the operation of superconducting quantum interference (SQUID) devices. The pulse-tube cooler (PTC) has the advantage of having no moving parts inside the cold head, reducing the vibrational noise and EMI generated. Two-stage Gifford-McMahon type pulse-tube coolers are able to reach temperatures below 4 K.[2] Many studies have investigated the use of pulse-tube coolers to cool high-$T_c$ SQUIDs.[3,4,5] Pulse-tube coolers are frequently used to cool low-$T_c$ SQUIDs in the field of millimeter-wave astronomy.[6]

Detectors for millimeter-wave astronomy are predominantly bolometers, which measure the incident electromagnetic power via the amount of heat deposited on an absorber, typically using a resistive thermometer. Bolometers can achieve high sensitivity by reducing thermal background power loading and other sources of thermal noise by cooling the bolometer and its thermal bath to cryogenic temperatures.[7] The nature of cooling an array of detectors on a telescope installed in a remote location, as is necessary for many millimeter-wave experiments, favors using a cryocooler over liquid cryogens if possible, and several experiments are now using pulse-tube coolers.[8,9] The current state-of-the-art millimeter-wave detectors are superconducting transition edge sensor (TES) bolometers, read-out by SQUID amplifiers. These bolometers achieve their extreme sensitivity to incident power by means of a superconducting film operating within its steep normal-to-superconducting resistance transition ($T_c \sim 0.5$ K).[10] When operated with a bath temperature $T < 0.25$ K, the inherent thermal carrier noise of the bolometer is minimized to the point that, after reducing read-out and other external noise sources, the detectors can reach a sensitivity where they are limited only by the photon noise of the thermal background they observe.[11] In this work we report on the initial design, testing, and operation of a pulse-tube cooled cryostat for TES bolometer testing at sub-Kelvin temperatures.

## CRYOGENICS

### Refrigeration

**Pulse-tube cooler**. The main cryocoolers used for this work was the PT415 cryorefrigerator by Cryomech Inc., a two-stage pulse-tube cooler with a base temperature of 2.8 K.[12] The remote motor option was chosen, as the motor is known to be a major source of noise. The motor was mounted adjacent to the cryostat, on the side of the cryostat's supporting structure, with vibrational and electrical isolation. A Cryomech vibration isolating bellows assembly was placed between the top of the pulse tube head and the cryostat. The PTC's first stage has 40 W cooling capacity at $T = 45$ K, and an unloaded base temperature of 32 K. The second stage has 1.5 W of cooling power at $T = 4.2$ K, and an unloaded base temperature of 2.8 K. The PTC and its associated compressor require only electricity and cooling water to operate.

**Evaporation refrigerator.** To achieve sub-Kelvin temperatures, the pulse-tube cooler is used in combination with a closed-cycle $^3$He/$^4$He/$^3$He evaporation refrigerator from Chase Cryogenics ('He-10 fridge').[13] This refrigerator has two cold heads, the "ultra-cold head" and "intermediate-cold head," as well as an additional heat sink point at the heat exchanger between the two heads. The "ultra-cold head" is expected to operate at approximately 250 mK under a typical load of 2 µW, with an unloaded base temperature of 220 mK. The "intermediate-cold heat" is expected to operate at approximately 350 mK with 50 µW of loading, with an unloaded base temperature of 330 mK. The refrigerator is a closed-cycle system: recycling is accomplished using entirely electrical controls. Because of heat generated during the cycle, the refrigerator requires at least 0.5 W of cooling power available. Voltage is initially applied to heaters at the helium cryopumps to liberate the gases. $^4$He is able to condense at a lower point cooled to the temperature of the 4 K stage, which is less than the critical temperature of $^4$He, $T = 5.19$ K. A voltage-controlled gas-gap heat switch is opene, and evaporative cooling of the He$^4$ lowers the temperature of the cold heads to approximately $T = 1$ K, which allows $^3$He to condense (critical temperature 3.35 K). Evaporative cooling of the two stages of $^3$He, with the intermediate-cold head acting as a buffer, results in the ultra-cold head reaching its base temperature of about 220 mK. The entire cycle can be accomplished in approximately 2 hours, and the hold time with typical loading is several days.

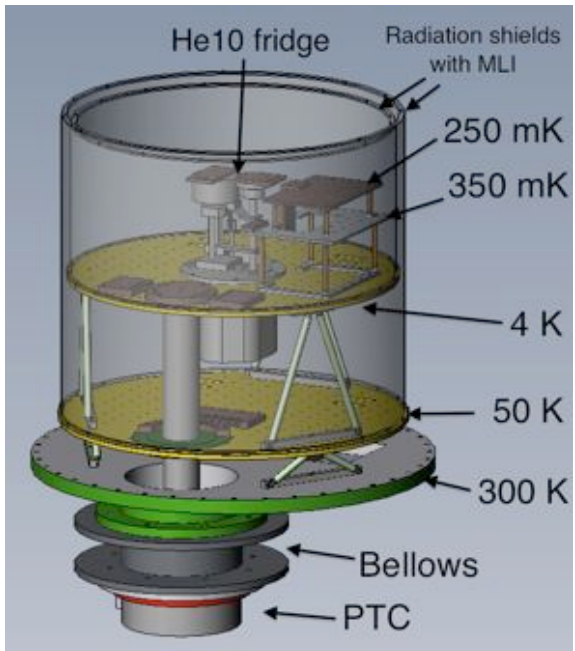

**Figure 1.** Solidworks model of cryostat, with outer shell hidden.

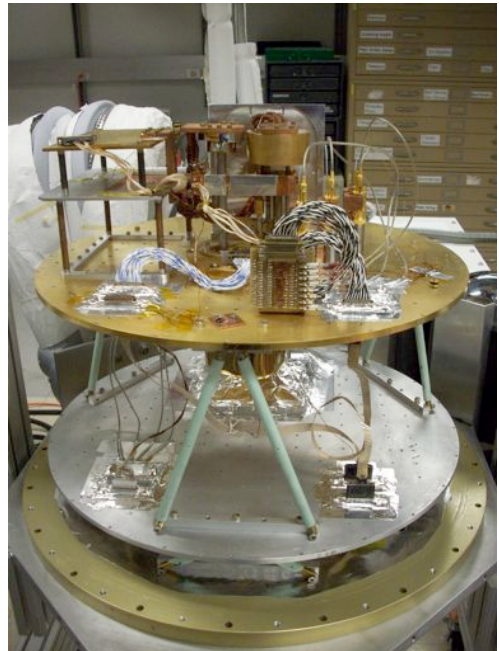

**Figure 2.** Photograph of interior of cryostat, rotated 180 degrees from Solidworks view

**Thermal architecture**

For convenience, the cold stages are referred to by their nominal temperature, e.g. "4 K stage." The cryostat architecture is shown in a Solidworks model in Figure 1, and a photograph is shown in Figure 2. The cold stages are supported by thermally isolating supports made of materials selected with a low thermal conductivity at the relevant temperature values, based on measurements made by Runyan and Jones.[14] For supports used at temperatures above 4 K, hollow rods of G-10 fiberglass-epoxy laminate were used. For the sub-Kelvin stages, conductive loading needs to be minimized as much as possible, and so materials with lower thermal conductivities were required. For these stages, hollow rods made of two types of Vespel® polyimide resin were used for supports, with thermal conductivities at 4 K approximately an order of magnitude lower than G-10, but with much higher cost. Vespel SP1 was used for the supports from 4 K to 350 mK. For temperatures below 2 K, the same polyimide resin with 40% graphite added, Vespel SP22, has an even lower thermal conductivity. This Vespel SP22 was used for supports from 350mK to 250 mK. All cryogenic wiring is small gauge wire made of low thermal conductivity material (manganin, phosphor bronze, superconducting NbTi for T < 10 K) with thermal breaks at each cold stage. A total of about 150 wires were permanently installed from room temperature to the 4 K stage, and as many as 48 wires can be brought to the 250 mK stage.

To thermally connect the refrigerators to the cold stages, heat straps were designed that would have a high thermal conductivity with minimal vibrational coupling. Copper speaker wire is ideal for this purpose as it is flexible and readily available in high purities. The thermal conductivity of copper depends strongly on purity at low temperatures, so speaker wire was selected (Bell'O SP7605 High Performance 14 AWG speaker wire) that was made of 99.99% oxygen-free copper, the highest purity readily available for commercial speaker wire. The heat straps consist of speaker wire stranded together and TIG welded to copper blocks to interface with the cold heads and the cold stages (see Figure 3). This general design was used for all the heat straps in the system.

**300 K stage.** The outer room temperature shell (Figure 4) is a large aluminum vacuum chamber capable of maintaining a vacuum of $P < 10^{-6}$ Torr. A turbomolecular pump with backing pump is used to pump down to $P = 10^{-3}$ Torr, before beginning cooling with the PTC. The vacuum pressure is further improved by cryopumping, reaching a pressure at the cryostat inlet of $P = 10^{-6}$ Torr. To improve the effectiveness of cryompumping, pieces of activated charcoal attached to a copper mount are mounted on the 4 K and 50 K stages. These "getters" are baked at 100° C between each run.

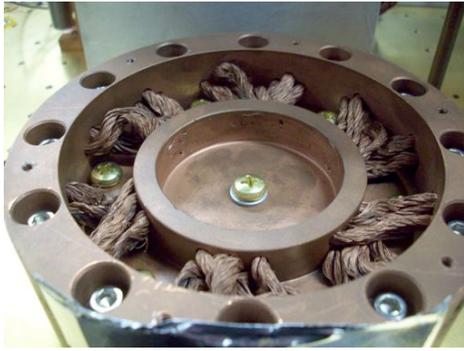

**Figure 3.** Heat strap to PTC at 4 K stage

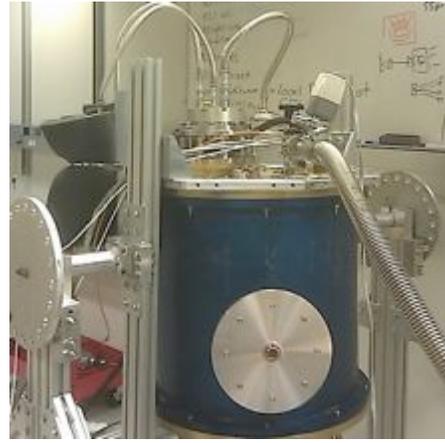

**Figure 4.** Outer shell vacuum chamber

**50 K stage.** The first stage of the PTC is used to cool an intermediate "50 K stage," with a large aluminum cold plate. This stage acts as a thermal break for radiative and conductive loading, and typical and acceptable loading can be near the capacity of the cooler, resulting in a temperature of about 40-45 K. A cylindrical aluminum radiation shield attaches to the 50 K cold plate and encloses the inner stages, with six layers of aluminized mylar (multi-layer insulation, MLI) to decrease emissivity and radiative loading. Aluminum tape is used to seal any gaps or holes that would result in additional radiative loading reaching the inner stages.

**4 K stage.** The second stage of the PTC is used to cool the main cold stage, the "4 K stage." This stage consists of a large plate made of oxygen-free high-conductivity (OFHC) copper, gold-plated for improved thermal contact. This stage is also used as a thermal break for radiative and conductive loading on the sub-Kelvin stages. A similar aluminum radiation shield is used with 10 layers of MLI, and any gaps here are also covered with aluminum tape. With 0 W loading, this stage is expected to reach 2.8 K, and the typical temperature achieved in our system is 2.9 K. The DC SQUIDs are mounted on this stage (operation requires T < 7 K), as well as readout circuits and the He-10 fridge.

**Sub-Kelvin stages.** The He-10 fridge has two cold heads as well as an additional heat sink point at the heat exchanger between the two heads, at about 1.5 K. The heat exchanger is used as heat sink for wires running to the sub-Kelvin stages. The "intermediate-cold head" is attached via heat strap to the "350 mK stage," an aluminum stage used for heat sinking wires as well as additional higher-temperature testing space. The "ultra-cold head" is attached to the 250 mK stage, a 6"x4" gold-plated OFHC plate used as the main cold stage for superconducting detector testing.

**READOUT SYSTEM**

**Dc SQUIDs**

A superconducting quantum interference device (SQUID) is a very sensitive magnetometer, with operation based on the principles of flux quantization and Josephson tunneling. A dc SQUID consists of two Josephson junctions in parallel in a superconducting loop which, when biased with an appropriate current, results in a voltage-to-flux V- $\phi$ relation that is periodic with applied flux. DC SQUIDs are typically operated in a flux-locked feedback loop, operating at an optimum working point located near the steepest part of the V- $\phi$ response. This linearizes the SQUID response and allows detection of very small changes in flux.[15]

Commercial low-$T_c$ dc SQUIDs were chosen for our readout system, a Quantum Design Model 5000 dc SQUID controller with four channels of Model 50 DC thin film SQUID sensors.[16] These SQUID sensors are made of a niobium/aluminum trilayer, with an operating temperature of T < 7 K, and an expected flux noise of < $5 \times 10^{-6}$ $\phi_0/\sqrt{Hz}$, where $\phi_0$ is the magnetic flux quantum, $\phi_0 = h/2e = 2.067 \times 10^{-15}$ Wb.

**Detector biasing and readout.**

In order to accurately measure the resistance of the TES detectors, both the voltage and current at the detector must be accurately measured. The detector current readout requires an amplifier with low input impedance (since $R_{TES}$<1 Ohm), high sensitivity, and high gain.[17] To keep readout noise sub-dominant to other noise sources, the current noise must be less than about 10 pA/$\sqrt{Hz}$. SQUID amplifiers are ideal candidates. The SQUID is connected in series with the detectors to act as a sensitive ammeter, with the current proportional to a voltage read out with the room temperature SQUID controller. Superconducting NbTi wire is used for all connections from the circuit board at 4 K to the detector and the SQUID sensor, eliminating any lead resistance. TES detectors also require a voltage bias to stably operate within the narrow superconducting transition using electrothermal feedback.[18] The custom detector voltage bias circuit, shown in Figure 5, consists of a current-biased shunt resistor at 4 K with R << $R_{TES}$, in parallel with the superconducting loop containing the detector in series with the input coil to the SQUID. The current in the entire loop is determined by measuring the voltage across a precision resistor at room temperature, and from this the voltage across the detector can be accurately determined. This voltage is measured using an instrumentation amplifier (Stanford Research Systems Model SIM911)[19] and a data acquisition device (DAQ) (Labjack UE9)[20]. The SQUID controller voltage is recorded with the same DAQ.

**Auxiliary inputs and readout**

In addition to the detector readout, additional inputs and readouts are brought into the cryostat for monitoring temperatures, controlling heaters, operating and cycling the He-10 fridge, and other purposes. Most of these were read out and controlled using off-the-shelf electronics, including many modules in a single mainframe from the SRS SIM series,[21] and various dc power supplies. The He-10 fridge cycle is controlled using custom Labview programs with serial readout of temperatures from the SIM mainframe and power supplies controlled via GPIB.

**Noise reduction techniques**

Much effort was necessary to reduce noise in the readout system and the surrounding environment. SQUIDs are sensitive to EMI, and the pulse tube cooler can be major contributor.[22] Other major electrical noise sources included environmental rf sources, the auxiliary readout and control systems, AC power mains in the building, and the vacuum system.

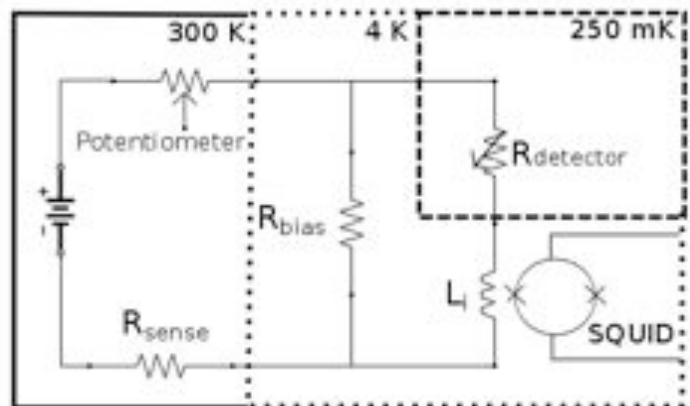

**Figure 5**. Detector voltage bias circuit diagram

Noise and unwanted ground connections from the vacuum system were eliminated, as cryopumping dominated over the effectiveness of the turbomolecular pump. The cryostat could be removed from the vacuum system once it reached its base temperature. Auxiliary readout noise and interference could be temporarily eliminated during the most sensitive science measurements by disconnecting everything but the necessary signal lines. The other sources of noise were not able to be eliminated entirely, but were mitigated to the point that they were negligible. Direct EMI from the PTC motor driver was reduced by mounting the motor as far from the cryostat case as possible (~2 ft away), as well as running the motor cable as far from other cables as possible. The readout lines were physically separated into three cables for the "clean" signal lines, He-10 fridge readout, and all other auxiliary readout. The signal cable runs separately from the other lines both inside and outside of the cryostat. All lines are filtered at the input to the cryostat, with pi filters (Spectrum Control 5000 pF Pi type filters)[23] inside rf sealed boxes at 300 K. All external cables were completely shielded, and sensitive signals were placed on individually shielded twisted pairs within the shielded cable. These inner shields were originally telescoping shields, floated at the cryostat end to prevent stray currents in the shield inducing low frequency noise on the twisted pair within. As testing progressed, it was found that the noise was dominated by rf noise, not low frequency noise, and keeping both ends of the shields at common potential improved performance. The final configuration with the best performance has the cryostat case, both ends of all shields, and "common ground" signals all kept at a common potential, with any stray ground connections and ground loops eliminated. Additional braided shield ground straps were added between the cryostat case and the common potential as an rf ground conductor. All AC mains connected equipment (power supplies, SQUID controller, computer, etc) is served from a common power phase.

## RESULTS

### Thermal Performance

Pulse-tube coolers can offer very good long-term stability compared to liquid cryogens, since they can run indefinitely. However, the main drawback can be that the pulsing action also causes low frequency temperature variations near the pulse-tube frequency. Changes in temperature change the effective area of a SQUID, which with a constant background magnetic field results in a change in flux. Calculations of expected noise contributions from temperature fluctuations for low-$T_c$ dc SQUIDs near 4 K predicts fluctuations of 0.1 K to have a noise contribution of $5 \times 10^{-7}$ $\varphi_0/\sqrt{Hz}$.[24] Additional thermal mass or thermal isolation from the cold head can mitigate fluctuations. These must be balanced with the limited thermal conductivity, which can result in long cool down times for large systems. For our system, cooling from room temperature to 4 K takes about 24-32 hours (Figure 6). At this point the He-10 fridge can be cycled and stable base temperatures are achieved after several hours. Temperature fluctuations from the pulse-tube cooler of 0.016 K with a period of 6 seconds are seen at the 4 K stage (Figure 7). These fluctuations do not affect the sub-Kelvin stages temperature, and the 250 mK stage temperature is stable to within approximately 0.001 K.

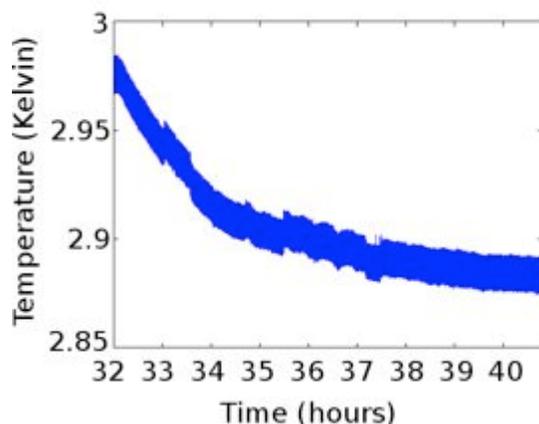

**Figure 6.** Typical cooldown vs. time, starting at room temperature at t=0.

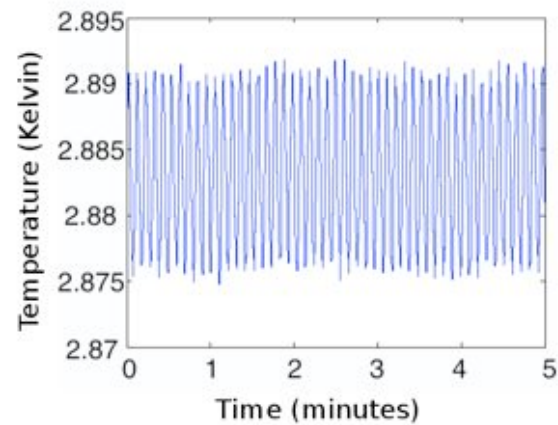

**Figure 7.** Temperature at 4 K stage after stabilization, showing small-scale temperature fluctuations from the pulse-tube head.

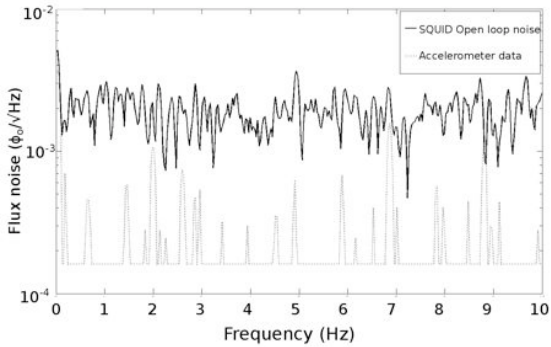 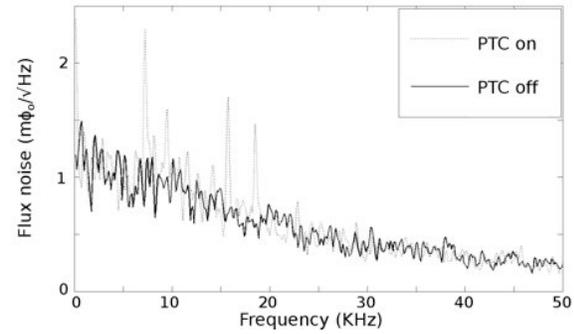

**Figure 8.** Low frequency SQUID noise.

**Figure 9.** High frequency noise, showing the peaks from the PTC's pulse-width modulated motor driver.

**Noise performance**

The white and 1/f noise of the SQUIDs was measured as described in the SQUID manual, and compared with the SQUID sensor specifications. With the SQUID in open loop operation, the output is a sensitive measure of noise in the system, and potential noise sources can be investigated and quantified. A spectrum analyzer (Stanford Research Systems Model SR770) was used to measure the noise spectral density of the zeroed open-loop voltage in many configurations.

Noise contributions at different frequencies were investigated. A unixaxial accelerometer (Endevco Corp. Model 2215, uniaxial accelerometer) on the outer case of the cryostat was used to investigate the spectrum of vibration peaks, to check for correlations with low frequency noise (Figure 8). As the accelerometer was only measuring on the outer case, it is only a relative measurement of the vibration amplitude at the cold stages, and there is always the possibility of resonances of internal structures resulting in a different vibration spectrum at the cold stages. A more thorough investigation would involve measuring vibrations directly at the cold stage. A

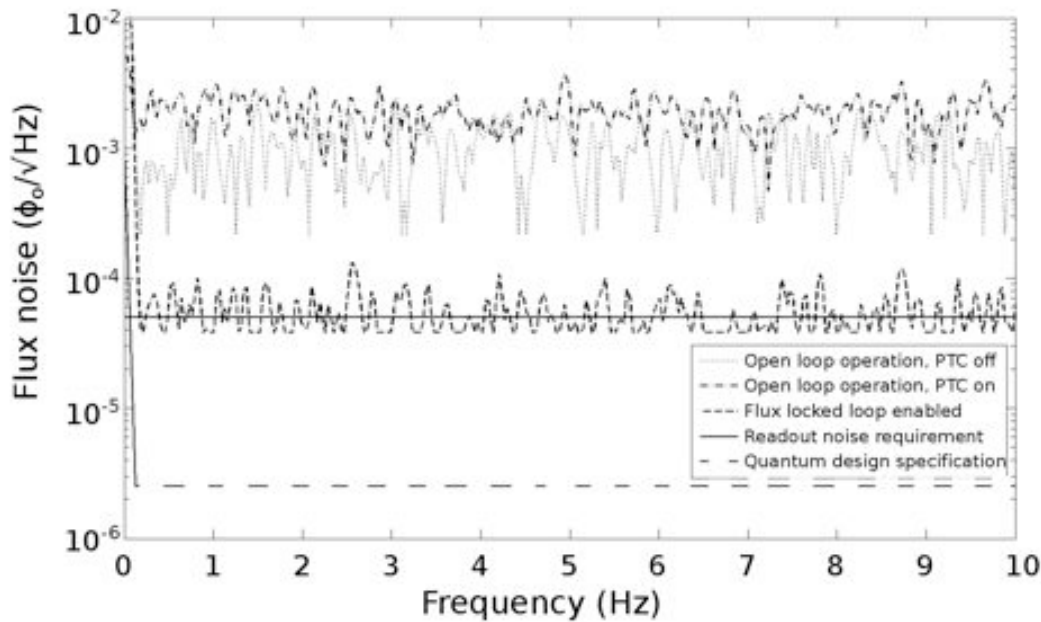

**Figure 10.** Flux noise performance of a single squid channel, compared to Quantum Design specifications.

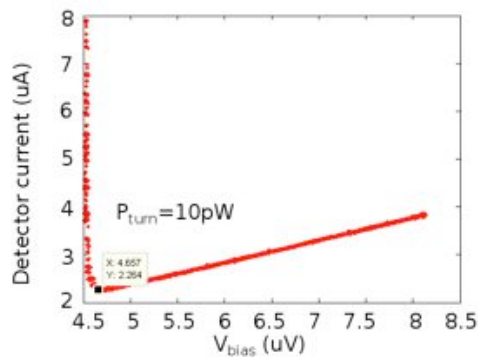
**Figure 11.** IV curve of TES bolometer, taken at T = 300 mK.

detailed study of the vibration spectrum of a pulse-tube cooler has been done, finding vibrations up to 15 KHz.[25] Temperature fluctuations, with a low amplitude and frequency, had no significant effect on noise. At high frequencies, peaks from the pulse-width modulated motor driver of the PTC dominate the noise (Figure 9). Replacing this with a low-noise linear driver could make significant improvements.[26]

Noise spectral densities of flux noise for a single SQUID channel are shown in Figure 10, comparing the baseline from Quantum Design with our achieved values. The baseline value from Quantum Design ($3\times10^{-6}$ $\varphi_0/\sqrt{Hz}$) is shown in red. Noise peaks from the PTC are apparent at lower frequencies (black) when compared to the data taken with the PTC off (blue). The noise floor is lowered and several peaks are eliminated in flux-locked loop operating mode (green).

Converting the flux noise seen up to 10 Hz to a current noise results in a current noise of approximately 10 pA/$\sqrt{Hz}$, which just meets our readout noise requirement of 10 pA/$\sqrt{Hz}$, based on having the readout noise be subdominant to other noise sources. The 1/f knee is approximately 0.2 Hz. At higher frequencies, the noise increases to a level of approximately 100 pA/$\sqrt{Hz}$. Since this is a test cryostat for detector prototyping and not long-term science observations, these noise levels are acceptable to operate and test detectors, but we will continue to work on improvements. Since high frequency noise dominates, for dc operation, further noise reduction is possible with low-pass filtering, and the noise does not affect our science results. Figure 11 shows a typical bolometer I-V curve, with a 1 KHz Butterworth low-pass filter on the SQUID output.

## CONCLUSIONS

We have demonstrated the successful design and operation of a pulse-tube cooled test cryostat with low-$T_c$ dc SQUID amplifiers. We found that high frequency noise from the pulse-tube cooler, especially its pulse-width modulated motor driver, dominated the noise but this noise can be mitigated to levels where SQUID operation and detector testing was uncompromised This high frequency noise could become an issue for our system in the future, and improvements will be made to increase rf shielding and reduce EMI interference from the pulse-tube cooler.

## ACKNOWLEDGMENTS

This work was supported by the Ax Center for Experimental Cosmology at the University of California, San Diego.

## REFERENCES

1. Radebaugh, Ray, "Pulse Tube Cryocoolers for Cooling Infrared Sensors," Proc. SPIE 4130: 363-379 (2000).

2. Radebaugh, Ray, "Development of the Pulse Tube Refrigerator as an Efficient and Reliable Cryocooler," Proc. Institute of Refrigeration (1999).
3. Lienerth, C, G Thummes, and C Heiden, "Progress in low noise cooling performance of a pulse-tube cooler for HT-SQUID operation," IEEE Transactions on Superconductivity, Vol 11, No. 1 (2001).
4. A. Rijpma et al, "Interference characterization of cryocoolers for a high-Tc SQUID-based fetal heart monitor," Cryocoolers 11: 793-802 (2002).
5. R. Hohman et al, "Comparison of low noise cooling performance of a Joule-Thomson cooler and a pulse-tube cooler using a HT SQUID," IEEE Transactions on Applied Superconductivity 9: 3688-3691 (1999).
6. Keating et al, "Ultra High Energy Cosmology with POLARBEAR," Proc. APS DPF (2011).
7. P. L. Richards, "Bolometers for infrared and millimeter waves," Journal of Applied Physics 76, 1-36 (1994).
8. SPT Collaboration: J. E. Ruhl et al., "The South Pole Telescope," Proc. SPIE 5498:11, 2004
9. K. Arnold et al., "The POLARBEAR CMB Polarization Experiment," Proc. of SPIE Vol. 7741, 7741E (2010).
10. K. D. Irwin and G. C. Hilton, "Transition-edge sensors," Topics Appl. Phys. 99, 63-149 (2005).
11. K. Arnold et al. 2010.
12. Cryomech, Inc., 113 Falso Dr., Syracuse, NY 13211 USA, www.cryomech.com
13. Chase Cryogenics Ltd., 140 Manchester Rd, Sheffield, UK, www.chasecryogenic.com
14. Runyan and Jones, "Thermal conductivity of thermally-isolating polymeric and composite structural support materials between 0.3 and 4K," *Cryogenics 48* 448-454 (2008).
15. J. Clarke and A. I. Braginski, SQUID Handbook, 2004 WILEY-CVH Verglag GmbH & Co.
16. Quantum Design, Inc., 6325 Lusk Blvd., San Diego, CA 92121 USA, www.qdusa.com
17. Irwin and Hilton 2005.
18. Irwin and Hilton 2005.
19. Stanford Research Systems, Inc. 1290-D Reamwood Ave., Sunnyvale, CA 94098 USA. www.thinksrs.com
20. Labjack Corporation, 3232 S Vance St STE 100, Lakewood, CO 80227 USA, www.labjack.com
21. Stanford Research Systems.
22. SQUID Handbook. J Clarke and A. I. Braginski. 2004 WILEY-CVH Verglag GmbH & Co.
23. Spectrum Control Inc., 8031 Avonia Rd., Fairview, PA 16415 USA
24. Schone, M. Muck, G. Thummes, and C. Heiden. "Investigation of the response of superconducting quantum interference devices to temperature variation." Rev. Sci Instr. 68 85-88
25. Chijioke, Akobuije, and John Lawall, "Vibration spectrum of a pulse-tube cryostat from 1Hz to 20kHz," Cryogenics 50 (4) (April): 266-270 (2010).
26. Zigmund Kermish, University of California, Berkeley, personal communication.